\documentclass[pra, twocolumn]{revtex4-1}

\usepackage{float}
\usepackage{verbatim}
\usepackage[utf8]{inputenc}
\usepackage[T1]{fontenc}
\usepackage[english]{babel}
\usepackage{ dsfont }
\usepackage{amsmath}
\usepackage{amssymb}
\usepackage{mathrsfs}
\usepackage{amsthm}
\usepackage{mathtools}
\usepackage{wasysym}
\usepackage{ulem}

\usepackage{makecell}

\usepackage{stfloats}

\newtheorem{definition}{Definition}[]

\usepackage{graphicx}
\usepackage{tikz}

\usepackage{hyperref}
\usepackage{framed}

\def\diracspacing{0.7pt}

\newcommand{\ket}[1]{| \hspace{\diracspacing} #1 \rangle} 
\newcommand{\ketbra}[2]{| \hspace{\diracspacing} #1 \rangle \langle #2 \hspace{\diracspacing} |} 

\newcommand{\norm}[1]{\|#1 \|}

\newcommand{\id}{\mathds{1}}

\newtheorem{Rmk}{Remark}
\DeclareMathOperator{\tr}{tr}
\newcommand{\gm}[1]{\textcolor{black}{#1}}

\newcommand{\de}[1]{\left(#1\right)}

\newcommand{\DE}[1]{\left\{{#1}\right\}}

\usepackage{natbib}
\begin{document}

\title{Quantum Conference Key Agreement: A Review}

\author{Gl\'{a}ucia Murta}
\author{Federico Grasselli}
\author{Hermann Kampermann}
\author{Dagmar Bru\ss}
\affiliation{Institut f\"ur Theoretische Physik III, Heinrich-Heine-Universit\"at D\"usseldorf, Universit\"atsstraße 1, D-40225 D\"usseldorf, Germany}

\begin{abstract}
\noindent Conference key agreement (CKA), or multipartite key distribution, is a cryptographic task where more than two parties wish to establish a common secret key.
A composition of bipartite quantum key distribution protocols can accomplish this task. However,
the existence of multipartite quantum correlations allows for new and potentially more efficient protocols, to be applied in future quantum networks.
Here, we review the existing quantum CKA protocols based on multipartite entanglement, both
in the device-dependent and the device-independent scenario. 
\end{abstract}

\maketitle

\section{Introduction}
\noindent Quantum mechanics can bring unprecedented advantages to the realization of information processing tasks. A remarkable example is quantum key distribution (QKD)\cite{BB84, E91}, arguably the most mature quantum technology. QKD allows two parties, Alice and Bob, to securely communicate by establishing a secret key that is information theoretically secure. {Security proofs are given for different levels of assumptions. In the scenario where the devices and/or quantum states are characterized}, robust security is proven for realistic parameters \cite{TLGR12,HT12} (see also \cite{TL17}) with implementations achieving long distances \cite{Hiskett06,Korzh15,Frohlich17}. {Also f}or the device-independent scenario, {i.e.\ no assumptions on the quantum states and on the working behavior of the devices,} a security proof in the fully adversarial scenario is well established \cite{EAT} \gm{(barring composability in case the devices are re-used, see Remark \ref{remarkDI})}. The required experimental parameters are characterized \cite{Murta19} for protocols based on the simplest Bell inequality \cite{CHSH}.

The extensive development of quantum technological applications allows near future applications which are based on genuine multipartite quantum protocols using shared multipartite entangled states in network structures \cite{Eppingnet1,Eppingnet2,Hahn2019,Pirker2018,lightmatter1,lightmatter2,satellite3}. Applications range from distributed quantum computing to genuine multipartite quantum communication protocols
which may lead to the quantum internet~\cite{Kimble2008,WEH18}.

{Here we focus on}
conference key agreement (CKA), {or} multiparty key distribution, {which} is a generalization of the task of key distribution to the scenario in which $N$ users wish to establish a common secret  key.
{This} allows the users to broadcast secure messages in a network. CKA can {e.g.} be achieved by, first establishing bipartite keys between  the users, {followed by securely} distributing a common key to all other users {via the bipartite keys}. This solution has been discussed to be inefficient in the classical scenario, and several classical protocols allowing the parties to establish a common key  were proposed (see e.g. \cite{CKA_classic_x,CKA_classic_y} and \cite{CKA_classic_a,CKA_classic_b}). In the quantum scenario, that is, when the parties can use quantum resources, a
secure conference key can also be established by using several bipartite QKD links.
{Bipartite quantum links are} already being implemented in small quantum networks
over metropolitan distances \gm{\cite{DARPA1,DARPA2,SECOQC,ChineseNet,SwissQuantum,TokyoNet,CambridgeNet}} and in larger networks spanning entire countries \cite{Beijing-ShanghaiLine1,SatelliteQKD1,SatelliteQKD2}.
{The long-term vision of a general quantum network, however, goes beyond mere bipartite links
and includes network
nodes that process quantum information, thus enabling the distribution of multipartite entangled
states across the network \cite{Eppingnet1}. In a  quantum network, quantum communication with
genuine multipartite entangled states may offer advantages over the bipartite case \cite{Epping},
and allow secure interactions between an arbitrary subset of the participating partners.}

{The rich structure of multipartite {entangled} quantum states opens the possibility
for a wide variety of new key distribution protocols. While}
protocols for CKA based merely on bipartite QKD do not bring much novelty in terms of the necessary quantum technologies or the theoretical tools required for the security analysis, {this changes when} protocols explor{e} multipartite entanglement. Here, quantum correlations can be exploited to devise truly multipartite schemes. This is the focus of this paper, namely we will review the proposals and developments regarding the use of multipartite quantum entanglement for the establishment of {a} conference key.

\section{Preliminaries}

\subsection{Multipartite entangled resources}

\noindent Multipartite quantum states have a more convoluted structure than the bipartite ones \cite{HHHH09,GT09,ES14,BZ16}.
Different classes of states can be defined according to their entanglement properties, and concepts such as \emph{$k$-separability} and \emph{genuine multipartite entanglement} arise (for a precise definition of these concepts, see \cite{HHHH09,BZ16,GT09}).
{For multipartite systems,  there exist different entanglement classes} that are not equivalent under stochastic local operations and classical communication (SLOCC) \cite{DVC00,dVSK13,BZ16}. In particular, in the tripartite case \cite{DVC00}, two nonequivalent classes of {genuinely multipartite} entangled states can be defined: the GHZ-class represented by the Greenberger–Horne–Zeilinger (GHZ) state \cite{GHZ}
\begin{align}
\label{ghz}
\ket{\rm GHZ}=\frac{1}{\sqrt{2}}\de{\ket{000}+\ket{111}},
\end{align}
and the W-class represented by the W state \cite{DVC00}
\begin{align}
\label{w}
\ket{\rm W}=\frac{1}{\sqrt{3}}\de{\ket{001}+\ket{010}+\ket{100}}.
\end{align}
These  classes of states also exhibit different physical properties. The GHZ-state is a direct generalization of Bell states to the multipartite case and maximally violates the well-studied family of $N$-party Bell inequalities called MABK \cite{MABK1,MABK2,MABK3}. However, the entanglement present in the GHZ-state is not robust to particle losses, while the W-state still exhibits bipartite entanglement when one particle is lost.

The 3-party GHZ and W states in Eqs.~(\ref{ghz}) and (\ref{w}) can be generalized in a straightforward way to $N$ parties. They constitute the resources for  quantum CKA protocols discussed in the following sections.

\subsection{Security}

\subsubsection{Security definition}

\noindent We consider  $N$ users, Alice, Bob$_1$, Bob$_2$, \ldots, Bob$_{N-1}$.
The users wish to establish a common string of bits that is unknown to any other party, in particular to any potential eavesdropper.

{The security of} a quantum conference key agreement protocol {is based on} two conditions: \emph{correctness} and \emph{secrecy}.

\begin{definition}[Correctness]
A CKA protocol is $\epsilon_{corr}$-correct if
\begin{align}
p(K_A=K_{B_1}=\ldots =K_{B_{N-1}})\geq 1-\epsilon_{corr},
\end{align}
where $K_A$, $K_{B_i}$ are the final keys held by Alice and Bob$_i$
{and $p(K_A=K_{B_1}=\ldots =K_{B_{N-1}})$ is the probability that all final
keys are identical.}
\end{definition}

\begin{definition}[Secrecy]
A CKA protocol is $\epsilon_{sec}$-secret if, for $\Omega$ being the event that the protocol does not abort,
\begin{align}
    p(\Omega) \frac{1}{2}\norm{\rho_{K_AE|\Omega}-\tau_{K_A}\otimes\rho_{E|\Omega}}\leq \epsilon_{sec},
\end{align}
where $p(\Omega)$ is the probability of the event $\Omega$, \gm{$\rho_{K_AE|\Omega}$ is the state shared by Alice and Eve at the end of the protocol given the event $\Omega$},  $\tau_{K_A}=\frac{1}{|S|}\sum_{s_i\in S}\ketbra{s_i}{s_i}$ is the maximally mixed state over all possible values that the key $K_A$ can assume, and $S=\DE{0,1}^{\times \ell}$ where $\ell$ is the length of the key $K_A$.
\end{definition}

Correctness implies that, at the end of the protocol, Alice and the Bobs share the same string of bits except for probability at most $\epsilon_{corr}$. The secrecy requirement states that Alice's key is randomly chosen among the set of possible strings and the eavesdropper has no information about the key, except for probability at most $\epsilon_{sec}$.
If a CKA protocol is $\epsilon_{corr}$-correct and $\epsilon_{sec}$-secret, then it is said to be $\epsilon_{s}$-correct-and-secret for all $\epsilon_{s}\geq\epsilon_{corr}+\epsilon_{sec}$.

Additionally, a useful CKA protocol should have a robust honest implementation. This is captured by the concept of \emph{completeness}.
\begin{definition}[Completeness]\label{def:complete} A quantum CKA protocol is $\epsilon_{c}$-complete if there exists an honest implementation of the protocol, such that the probability of not aborting is greater than $1-\epsilon_{c}$.
\end{definition}

Finally, the security of a quantum CKA protocol can be summarized as~\cite{PR14}:
\begin{definition}[Security of a quantum CKA protocol]\label{def:security} 
A quantum CKA protocol is ($\epsilon_s,\epsilon_{c}$)-secure if
\begin{itemize}
    \item[(I)] (Soundness) For any implementation of the protocol, it is $\epsilon_s$-correct-and-secret.
    \item[(II)] (Completeness)  There exists an honest implementation of the protocol, such that the probability of not aborting is greater than $1-\epsilon_{c}$.
\end{itemize}
\end{definition}

Definition~\ref{def:security} implies composable security~{\cite{Canetti01,BM04,PR14}}. This means that the conference key generated by a protocol satisfying the conditions stated in Definition \ref{def:security} is composable secure and therefore can be used as a building block for further protocols (this, however, cannot always be inferred in the device-independent scenario, see Remark~\ref{remarkDI} in Section~\ref{sec:DI}).

The quantum left-over hashing lemma~\cite{RennerThesis,TSSR11} establishes that a secret conference key can be obtained if the key length $\ell$ is slightly shorter than
\begin{align}
    \ell \lesssim H_{\min}^{\epsilon}(A_1^n|E)
\end{align}
where $H_{\min}^{\epsilon}(A_1^n|E)$ is the conditional smooth min-entropy \cite{TomamichelBook} evaluated for the classical-quantum (cq) state $\rho_{A_1^nE}$ composed of Alice's raw key of size $n$ and the quantum side information of a potential eavesdropper.

The conditional smooth min-entropy of a cq-state $\rho_{AE}$ is defined as
\begin{align}
	H_{\min}^{\epsilon}(A|E)=
	\sup_{\tilde{\rho}_{AE}\in \mathcal{B}^{\epsilon}(\rho_{AE})} H_{\min}(A|E),
	\end{align}
	where $\epsilon \in [0,1)$, and the supremum is taken over positive sub-normalized operators that are $\epsilon$-close to $\rho_{AE}$ in the purifying distance~\cite{TomamichelBook}.
	And the conditional min-entropy, $H_{\min}(A|E)$, of a classical variable $A$ conditioned on the quantum side information $E$ is closely related to the optimal probability of the eavesdropper guessing the value of $A$, $p_{guess}(A|E)$,  \cite{KRS09}
\begin{eqnarray}\label{eq:Hmin}
H_{\min}(A|E)=- \log p_{guess}(A|E).
\end{eqnarray}
For a precise definition and properties of entropic quantities we refer the reader to \cite{TomamichelBook}.

The main task in the security proof of a conference key agreement protocol is to estimate $H_{\min}^{\epsilon}(A_1^n|E)$. Note that this is very similar to the bipartite case of quantum key distribution. In fact, the secrecy condition only depends on the correlations between the eavesdropper and Alice's string. However, in the multipartite scenario the parties need to ensure that all of the Bobs correct their raw key so that the correctness requirement is satisfied.

\subsubsection{Security model}

\noindent In the scenario where $N$ parties wish to securely communicate, the adversary is an external party, Eve, who can eavesdrop on all the exchanged public communication. Moreover, Eve might try to tamper with the quantum channels and explore correlations with the generated conference key.

Similar to the bipartite case, we can also classify the attacks performed by the eavesdropper into three categories:
\begin{enumerate}
    \item \emph{Individual attacks}: the eavesdropper can only attack individually each round of the protocol. In this case she is assumed to have no quantum memory, and therefore her best strategy is to perform a measurement on her quantum  side information at each round.
    \item \emph{Collective attacks:} Eve is assumed to perform the same attack for each round of the protocol, that is, her {quantum side information is identically and independently distributed (IID) with respect to different rounds}. 
    Differently from individual attacks, Eve is now assumed to have a quantum memory. Therefore, she can store her quantum side information at each round and perform a global operation on it at the end of the execution of the protocol.
    \item \emph{Coherent attacks:} This is the most general type of attack where there are no assumptions on the capabilities of the eavesdropper, except that she is bounded by the laws of quantum mechanics. In this case the states shared by the parties at each round may have arbitrary correlations with previous and future rounds.
\end{enumerate}

\subsection{Generic Protocols}

\noindent The goal of quantum conference key agreement is that the $N$ users make use of their shared quantum resources together with local operations and public communication in order to establish a secure conference key.

In {the following} section we will present the proposed quantum protocols that perform the task of CKA, making use of multipartite entanglement. The protocols we will discuss consist of the following main steps:

\begin{enumerate}
    \item \textbf{Preparation and distribution:} A source distributes a multipartite entangled state to the $N$ parties. This step is repeated $n$ times.

    \item \textbf{Measurements:} Upon receiving the systems, the parties perform local measurements and record the classical outcome. The measurements are randomly chosen according to the specifications of the protocol. One of the possible measurement settings is used with higher probability and is called the \emph{key generation} measurement. The other measurements are used for \emph{test rounds}, which only occasionally occur. \gm{A short pre-shared key can be used to determine if a round is a key generation round or a test round. Alternatively, the parties can implement a sifting step \cite{TL17} to select rounds where the same type of measurements were performed. }

    \item \textbf{Parameter estimation:} The parties announce the inputs and outputs of their test rounds and of some {randomly chosen} key generation rounds {which are used}  to estimate their correlation and the potential influence of an eavesdropper.
    {At the end of this step each party is left with a string of $n_{raw}<n$ bits, which constitute their \emph{raw key}.}

    \item \textbf{Information reconciliation {(error correction)}:} The parties publicly exchange classical information in order for the Bobs to correct their raw keys 
    to match Alice's string. 
    In the multipartite case, the information reconciliation protocol needs to account for the correction of the strings of all the Bobs.

    \item \textbf{Privacy amplification:} Alice randomly picks a hash function, chosen among a two-universal family of hash functions (see \cite{RennerThesis}), and communicates it to the Bobs. Every party applies the hash function to turn her/his partially secure string of $n_{raw}$ bits into a secure key of $\ell< n_{raw}$ bits.
       \end{enumerate}

The key rate of a protocol is given by
\begin{align}
    r=\tau \frac{\ell}{n}
\end{align}
where $\tau$ is the repetition rate of the setup, i.e. the inverse of the time it takes to implement one round of preparation and measurement of the quantum systems. In the following sections we will typically take $\tau=1$ as we will not be focused on any specific experimental implementation. The key rate in the limit of infinitely many rounds, $n \rightarrow \infty$, is called the \emph{asymptotic key rate} and denoted $r_{\infty}$.

\section{Protocols for multi-qubit states}

\subsection{GHZ state protocols}\label{sec:NBB84}
\noindent
The first proposals of quantum conference key agreement protocols explore the multipartite correlations exhibited by the $N$-party GHZ state:
\begin{align}\label{GHZstate}
    \ket{{\rm GHZ}_N}=\frac{1}{\sqrt{2}}\de{\ket{00\ldots0}+\ket{11\ldots1}},
\end{align}
where $\DE{\ket{0},\ket{1}}$ is the $Z$-basis, composed by the eigenstates of the Pauli operator $\sigma_z$.
The GHZ state satisfies all the desired conditions for a conference key agreement protocol: the outcomes of measurements in the $Z$-basis are perfectly correlated, random and uniformly distributed. Interestingly, for $N\geq 3$ this perfect correlation can only be achieved if all the parties measure in the $Z$-basis. As shown in \cite{Epping}, even bipartite perfect correlation cannot be obtained if the parties choose a different basis. This represents a drastic difference from the bipartite case ($N=2$). Indeed, if Alice and Bob share the maximally entangled state $\ket{\Phi}=\frac{1}{\sqrt{2}}\de{\ket{00}+\ket{11}}$, for each choice of local basis for Alice, there exists a local basis for Bob such that their outcomes exhibit perfect correlation. This property is exploited in the bipartite six-state \cite{sixstate} and BB84 \cite{BB84} protocols for QKD.

Early proposals of protocols that employ the GHZ state to establish a conference key between three parties were presented in \cite{Cabello00}. 
Security is proved, against individual attacks, for the ideal case where Alice can prepare and distribute perfect GHZ states. Robustness to noise is not considered.
In Ref.~\cite{CL07}, Chen and Lo proved the security of quantum conference key agreement based on the distillation of GHZ states \cite{MS02,MPPVK98}. 
They derive distillation
rates for a protocol based on an improved version of the multi-party hashing method~\cite{MS02}. These 
 rates correspond  to conference key
rates, due to the fact that the multi-party hashing distillation protocol~\cite{MS02}
can be implemented by classical post-processing of the raw key.
Ref.~\cite{CL07} also considers distillation rates when recurrence protocols are applied before the multi-party hashing.
Recurrence protocols are based on CSS codes \cite{CS96,Steane96} and, if certain conditions are met, they can also be translated to a classical post-processing of the generated raw keys, in a similar fashion to the bipartite case \cite{GL03}. Ref.~\cite{CL07} modifies the recurrence protocol introduced in \cite{MPPVK98}, using ideas of \cite{GL03}, to design a protocol that can be converted to classical post-processing of the raw key.
This type of classical post-processing of the raw key  requires two-way communication and  was denoted advantage distillation \cite{Maurer93,BA07,KBR07}.

In the following subsections, we present specific protocols with GHZ states that can be regarded as the generalization of the six-state and the BB84 protocols to the multipartite case.

\subsubsection{Multiparty six-state protocol}\label{sec:Nsixstate}
\noindent
The quantum conference key agreement protocol introduced in \cite{Epping} can be seen as a generalization of the six-state QKD protocol \cite{sixstate} to the multipartite case. Indeed, in Ref.~\cite{Epping} the parties perform measurements in the three bases $\DE{X,Y,Z}$. Measurements in the $Z$-basis are used with higher frequency, and they constitute the key generation rounds. The $X$-basis and $Y$-basis are instead used in fewer rounds, specifically in the test rounds, in order to estimate the information available to a potential eavesdropper.

{From} the parameter estimation {rounds}, the statistics of the $Z$-measurements is used to estimate the qubit error rates (QBERs) and thus to determine the information that needs to be communicated by Alice for information reconciliation. The bipartite QBERs, $Q_{AB_i}$, for $1\leq i\leq N-1$, are the probabilities that the outcome of a $Z$-measurement by Bob$_i$ disagrees with Alice's $Z$-measurement outcome. In the multipartite scenario we can also define the total QBER $Q_Z$ as the probability that at least one Bob obtains an outcome different than Alice. If the $N$ parties share a state $\rho$, the QBER $Q_Z$ is given by
\begin{align}
    Q_Z=1-\tr \de{\rho\de{\ketbra{0}{0}^{\otimes N}+\ketbra{1}{1}^{\otimes N}}}.
\end{align}
With the statistics of the test rounds, the parties want to estimate the expected value of the operator $X^{\otimes N}$. Since the multipartite GHZ state does not exhibit perfect correlation in more than one basis \cite{Epping}, the QBER $Q_X$ is defined as the probability that the $X^{\otimes N}$-measurement gives a result that differs from the ideal case:
\begin{align}
    Q_X=\frac{1-\langle X^{\otimes N}\rangle}{2}. \label{QX}
\end{align}
Note that if the parties share the GHZ state \eqref{GHZstate}, then the corresponding $Q_X$ is zero.

A crucial step in the security analysis of the protocol presented in Ref. \cite{Epping} is a reduction to depolarized states. An $N$-qubit depolarized state is a state of the form:
\begin{align}\label{dep_state}
\begin{split}
    \rho_{\rm dep}=\lambda_{0,\vec{0}}&\ketbra{\psi_{0,\vec{0}}}{\psi_{0,\vec{0}}}+\lambda_{1,\vec{0}}\ketbra{\psi_{1,\vec{0}}}{\psi_{1,\vec{0}}}\\
    &+ \sum_{\sigma, \vec{u}\neq\vec{0}}\lambda_{\vec{u}}\ketbra{\psi_{\sigma,\vec{u}}}{\psi_{\sigma,\vec{u}}},
    \end{split}
\end{align}
where
\begin{align}
    \ket{\psi_{\sigma,\vec{u}}}=\frac{1}{\sqrt{2}}\de{\ket{0}\ket{\vec{u}}+(-1)^{\sigma}\ket{1}\ket{\vec{\bar{u}}}}
\end{align}
for $\vec{u}\in \DE{0,1}^{\times (N-1)}$, {$\vec{\bar{u}}=\vec{u}\oplus \vec 1$}, and $\sigma\in \{0,1\}$. The states $\DE{\ket{\psi_{\sigma,\vec{u}}}}_{\sigma,\vec{u}}$ form a basis, denoted as the GHZ basis.
The depolarized GHZ state is then diagonal in the GHZ basis and such that $\lambda_{0,\vec{u}}=\lambda_{1,\vec{u}}\equiv\lambda_{\vec{u}}$ for $\vec{u}\neq \vec{0}$.

For a state of the form \eqref{dep_state}, one finds that
\begin{align}
   Q_Z(\rho_{\rm dep})=1-(\lambda_{0,\vec{0}}+\lambda_{1,\vec{0}}),
\end{align}
and
\begin{align}
    Q_X(\rho_{\rm dep})=\frac{1-(\lambda_{0,\vec{0}}-\lambda_{1,\vec{0}})}{2}.
\end{align}

Finally, the asymptotic key rate for the depolarized state \eqref{dep_state} is given as a function of $Q_X$, $Q_Z$ and the bipartite QBERs $Q_{AB_i}$ \cite{Epping}:
\begin{align}
\begin{split}
    r_{\infty}=&(1-Q_Z)\de{1-\log (1-Q_Z)}\\
    &+\de{1-\frac{Q_Z}{2}-Q_X}\log\de{1-\frac{Q_Z}{2}-Q_X}\\
    &+\de{Q_X-\frac{Q_Z}{2}}\log\de{Q_X-\frac{Q_Z}{2}}\\
    &-\max_{1\leq i \leq N-1} h\de{Q_{AB_i}}
\end{split}.
\end{align}

For the generality of the security analysis of \cite{Epping}, it remains to argue that the reduction to depolarized states, \eqref{dep_state}, is not restrictive. Any $N$-qubit state can be brought to the form \eqref{dep_state} by successive application of the following set of local operations \cite{DCT99,DC00}
\begin{align}
\begin{split}
    \mathcal{D}=\DE{X^{\otimes N}}&\cup \DE{Z_{AB_j}|1\leq j\leq N-1}\\
    &\cup \DE{R_k|1\leq k\leq N-1}  \label{dep_op}
    \end{split},
\end{align}
where the operations $Z_{AB_j}$ and $R_k$ are defined as
\begin{align}
    Z_{AB_j}=Z_A\otimes Z_{B_j}\otimes I_{B_{[N-1]\setminus j}}
\end{align}
and
\begin{align}
    R_k={\rm diag}(1,i)_A\otimes {\rm diag}(1,-i)_{B_k}\otimes I_{B_{[N-1]\setminus k}}.
\end{align}

Indeed, the application of the map
\begin{align}\label{dep_map}
    \rho \mapsto \tilde{\rho}=\circ_{i=1}^{2N-1}\mathcal{D}_i[\rho]
\end{align}
where
\begin{align}
    \mathcal{D}_{i}[\rho]=\frac{1}{2}\rho+\frac{1}{2}D_i \rho D_i^{\dag}\,;\, D_i \in \mathcal{D},
\end{align}
brings any $N$-qubit state to the form \eqref{dep_state}.

A crucial observation is that the map \eqref{dep_map} can be implemented in the protocol by flipping the outcomes of some of the measurements and adding additional measurements in the $Y$-basis \cite{Epping}.

Consider first the set of operations $\DE{X^{\otimes N}}\cup \DE{Z_{AB_j}|1\leq j\leq N-1}$. Successive application of these operations brings any $N$-qubit state to the GHZ-diagonal form
\begin{align}
    \tilde{\rho}=\sum_{\sigma,\vec{u}}\lambda_{\sigma,\vec{u}}\ketbra{\psi_{\sigma,\vec{u}}}{\psi_{\sigma,\vec{u}}}.
\end{align}
For the key generation rounds, in which Alice and the Bobs measure in the $Z$-basis, the application of $Z_{AB_j}$ does not have any effect on the final outcomes and the operation $X^{\otimes N}$ can be equivalently applied by Alice and the Bobs by flipping their $Z$-measurement outcomes. For the estimation of $X^{\otimes N}$ in the test rounds, the operations $\DE{X^{\otimes N}}\cup \DE{Z_{AB_j}|1\leq j\leq N-1}$ have no effect, as can be seen by the fact that they commute with $X^{\otimes N}$.

The application of the operations $\DE{R_k}$ 
is what finally brings the state to the depolarized form \eqref{dep_state}. They have no effect on the key generation rounds as they do not change the outcome of the $Z$-measurements. For the test rounds, the action of  $R_k$ is more subtle. As shown in \cite{Epping}, the action of $R_k$ followed by a measurement in the $X$-basis is equivalently implemented by Bob$_k$ performing a $Y$-basis measurement. Therefore the action of the operators $\DE{R_k}$, which are essential to simplify the security analysis of the protocol introduced \cite{Epping}, can be implemented in the protocol by adding $Y$-basis measurements to the test rounds.

In Ref.~\cite{Epping} the authors show that in a quantum network with quantum routers, {for a bottleneck configuration with constrained channel capacity}, the multipartite six-state protocol based on the GHZ state leads to higher rates as compared to several implementations of bipartite QKD, when the gate quality is above certain threshold value.

A security analysis of the multiparty six-state protocol against coherent attacks taking into account finite size effects was presented in \cite{Grasselli_2018}. 

\subsubsection{Multiparty BB84 protocol}
\noindent
In Ref.~\cite{Grasselli_2018}, {also} a multipartite version of the BB84 protocol was introduced: here, the parties only need to perform measurements in two bases, the $Z$-basis and the $X$-basis. The security analysis is based on the uncertainty relation for smooth entropies \cite{TR11}. This technique has previously been used in the bipartite case \cite{TLGR12,TL17} for the security proof of the BB84 protocol in the finite regime for parameters that are compatible with current technology. The uncertainty relation establishes that for a pure state $\ket{\psi_{A\vec{B}E}}$, if Alice can perform measurements in two bases, say the $X$-basis and the $Z$-basis, then the following relation is satisfied:
\begin{align}
H_{\min}^{\epsilon}(Z_1^m|E)\leq q-H_{\max}^{\epsilon}(X_1^m|B_1\ldots B_{N-1})
\end{align}
where the {conditional smooth min-}entropy on the l.h.s. is evaluated for the cq-state shared by Alice and Eve when Alice measures her systems in the $Z$-basis, and the {conditional smooth max-}entropy on the r.h.s.\ is evaluated for the cq-state shared by Alice and the Bobs when Alice measures her systems in the $X$-basis. {For a precise definition of $H_{\max}^{\epsilon}$ we refer the reader to \cite{TomamichelBook}.} 
The term $q$ quantifies the incompatibility of the two measurements used by Alice, and for the case where Alice can measure $X$ or $Z$ the quality factor $q$ for the $m$ rounds will be equal to $m$.

The quantity $H_{\max}^{\epsilon}(X_1^m|B_1\ldots B_{N-1})$ can be estimated by using the $X$-measurements performed by the Bobs \eqref{QX}. Indeed, the data processing inequality guarantees that
\begin{align}
    H_{\max}^{\epsilon}(X_1^m|B_1\ldots B_{N-1})\leq H_{\max}^{\epsilon}(X_1^m|\vec{X}_1^m), \label{dataproc}
\end{align}
where $\vec{X}_1^m$ contains the $X$-outcomes of every Bob, had the Bobs measured in the $X$-basis in the $m$ rounds. Clearly the entropy on the r.h.s. of Eq.~\eqref{dataproc}, that is the entropy of Alice's $X$-outcome string given the $X$-outcome strings of the Bobs, can be estimated via the $X$-basis error defined in Eq.~\eqref{QX}.

Finally Ref. \cite{Grasselli_2018} establishes the asymptotic secret key rate of the multiparty BB84 protocol
\begin{align}
    r_{\infty}=1-h(Q_X)-\max_{1\leq i \leq N-1}h(Q_{AB_i}).
\end{align}

\subsubsection{Comparison of multiparty six-state and  BB84 protocols}
\begin{figure}[!h]
	\centering
	\includegraphics[width=0.9\linewidth,keepaspectratio]{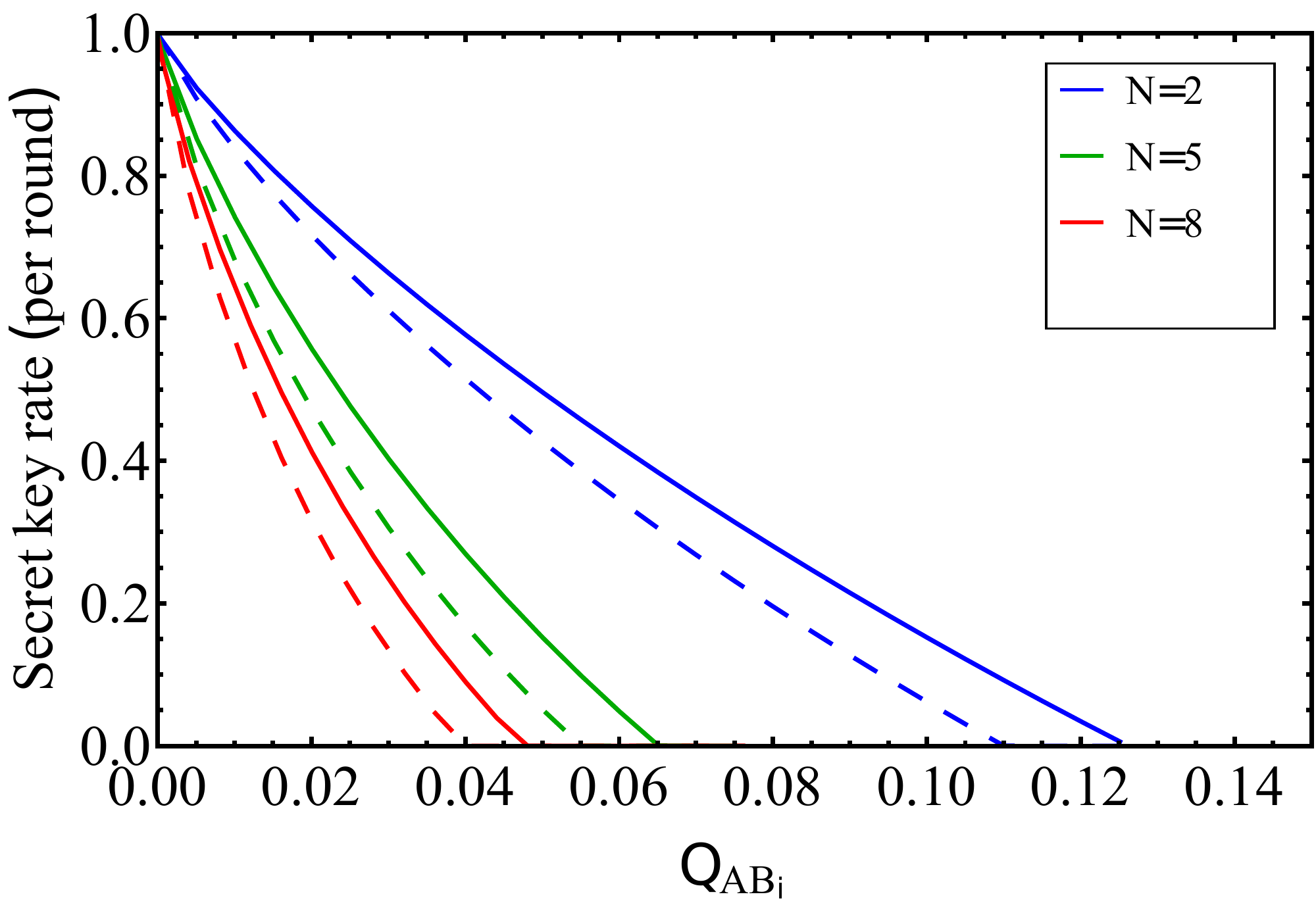}
	\caption{Asymptotic secret key rates of the multipartite six-state (solid) \cite{Epping} and BB84 (dashed) \cite{Grasselli_2018} protocols as a function of the bipartite QBER between Alice and any Bob, for a local depolarizing noise model. The rates are plotted for different numbers of parties ($N=2,5,8$, right to left). The plot shows that the multipartite six-state protocol asymptotically outperforms the multipartite BB84 protocol.}
	\label{asymptotickey-plot}
	\includegraphics[width=0.9\linewidth,keepaspectratio]{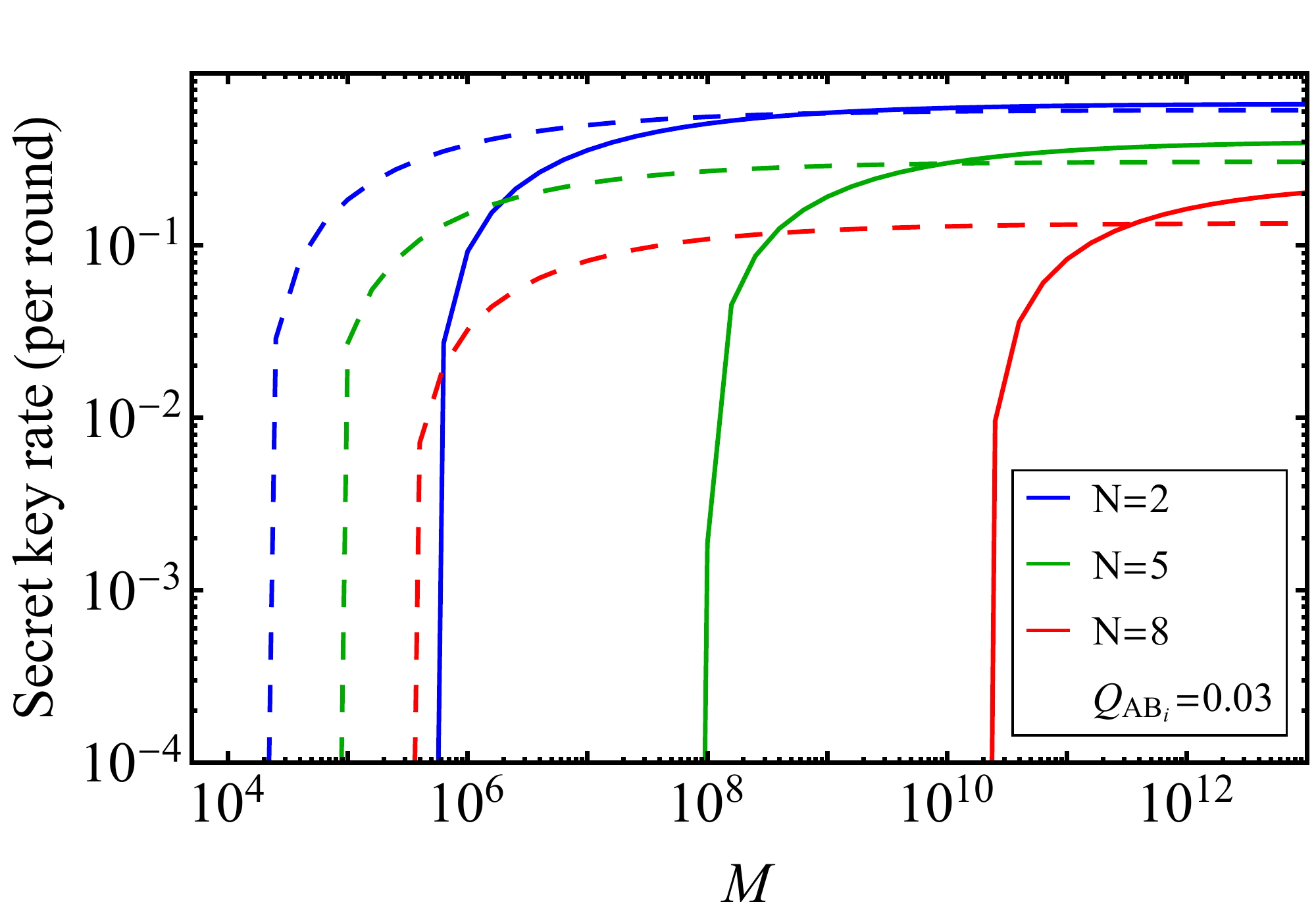}
	\caption{Secret key rates of the multipartite six-state (solid) \cite{Epping} and BB84 (dashed) \cite{Grasselli_2018} protocols as a function of the total number of rounds $M$, for different number of parties ($N=2,5$ and 8, left to right) and fixed bipartite QBER ($Q_{AB_i}=0.03$). The noise model employed is the local depolarizing channel given in Eqs.\eqref{depolarizedGHZ} and \eqref{localdep}. A non-null conference key can be obtained for fewer rounds with the multipartite BB84 protocol, compared to the multipartite six-state protocol, and the advantage of the former protocol increases with the number of parties.}
	\label{finitekey-plot}
\end{figure}
\noindent
For any specific implementation, the asymptotic key rates obtained by the multiparty six-state protocol \cite{Epping} are higher than those obtained by the multiparty BB84 \cite{Grasselli_2018}. This is because more structure can be ensured about the underlying state in the protocol presented in \cite{Epping}.
For instance, consider the implementation where Alice prepares a GHZ state and distributes it to each of the Bobs using a qubit depolarizing channel. The state shared by the parties is thus
\begin{align}
    \rho_{A\vec{B}}=\mathcal{D}_2^{\otimes (N-1)}\ketbra{{\rm GHZ}_N}{{\rm GHZ}_N}, \label{depolarizedGHZ}
\end{align}
where
\begin{align}
    \mathcal{D}_2(\rho)=(1-\nu)\rho+\nu\frac{\id}{2}. \label{localdep}
\end{align}
Figure~\ref{asymptotickey-plot} shows the comparison of the asymptotic key rates achieved by the two multiparty protocols ($N=2,5,8$) in the specific implementation given by the noise model in Eq.~\eqref{depolarizedGHZ}. The key rates are plotted as a function of the bipartite QBER between Alice and any Bob, which turns out to be a simple function of the noise parameter characterizing the depolarizing channel: $Q_{A B_i}=\nu/2$. The figure confirms that, asymptotically, the multipartite six-state protocol \cite{Epping} overcomes the multipartite BB84 \cite{Grasselli_2018} in terms of performance.

Ref.~\cite{Grasselli_2018} also performs a complete security analysis in the finite-key regime for the multiparty six-state and multiparty BB84 protocol.
Regarding the rates in the finite-key regime, {it was shown} that, even though the six-state protocol can tolerate higher noise, for the low-noise regime a non-zero conference key rate can be proven for the multiparty BB84 protocol using a significantly smaller number of rounds. This is confirmed by Figure~\ref{finitekey-plot}, where the secret key rates of both protocols are plotted as a function of the total number of protocol rounds, having fixed the bipartite QBER. The noise model employed is the same used for Fig.~\ref{asymptotickey-plot}, i.e. the local depolarizing channel given in Eq.~\eqref{localdep}. It is important to remark that the lower threshold on the minimum number of signals for a non-zero key by the multiparty BB84 protocol, may be simply due to the techniques used to compute the key rates. The finite-key rates of the multipartite six-state are derived using the post-selection technique \cite{CKR09} in combination with the finite version of the asymptotic equipartition property \cite{AEP} (see also \cite{RennerThesis}). These techniques might lead to higher overhead terms in the finite-key regime and therefore to a less tight estimate than what can be obtained using the uncertainty relation for smooth entropies \cite{TR11}. However, due to the fact that in the multiparty six-state protocol the parties are required to perform three distinct measurements, the uncertainty relation is not applicable.

\subsubsection{Prepare-and-measure implementation}
\noindent Even though entanglement plays an essential role for the security of bipartite QKD, it is known that some QKD protocols have a corresponding prepare-and-measure implementation that does not require any entanglement.
The BB84 protocol, for example, can be implemented with Alice transmitting single qubit states to Bob.

Similarly, in the multipartite case we can also talk about a corresponding prepare-and-measure implementation. However, now this reduction will require the preparation of some $(N-1)$-entangled states \cite{Epping}.

Indeed for the key generation rounds, in which the parties are performing measurements in the $Z$-basis, Alice could instead randomly choose her bit and prepare $(N-1)$ copies of the corresponding single qubit state to send to the Bobs, $\ket{0}^{\otimes (N-1)}$ or $\ket{1}^{\otimes (N-1)}$. Although entanglement is not required to reproduce the statistics of the key generation rounds, the corresponding state shared by the Bobs when Alice performs a measurement in the $X$-basis or $Y$-basis is entangled. Therefore, for the test rounds, Alice is required to prepare an $(N-1)$-entangled state.

For example, when Alice performs an $X$-measurement, given that she obtains the outcome $a$,  the corresponding state that she has to distribute to the Bobs is the $(N-1)$-entangled state:
\begin{align}
    \ket{\psi_a}_{B_1\ldots B_{N-1}}=\frac{1}{\sqrt{2}}\de{\ket{00\ldots0}+(-1)^a\ket{11\ldots1}}.
\end{align}

The prepare-and-measure equivalence significantly reduces the resources required for the implementation of the protocols \cite{Grasselli_2018,Epping}, as Alice needs to control $(N-1)$-partite entanglement instead of $N$-partite entanglement. This can have significant practical implications especially in the {\emph{noisy intermediate scale}} (NISQ) era \cite{NISQ}. Moreover, it is important to remark that, for most of the rounds, the key generation rounds, Alice can in fact prepare {product} states, and entanglement is only required in a small fraction of the rounds for the purpose of parameter estimation.

A prepare-and-measure protocol in which Alice only needs to send separable states was proved secure for the case $N=3$ in \cite{Matsumoto07}. However, {when extending the protocol to an arbitrary number of parties $N$ the states distributed by Alice would become {increasingly} distinguishable as $N$ increases, which would allow an eavesdropper to retrieve more information about the key, while causing less disturbance. Thus, the secret key rate would decrease with increasing $N$,
even for a perfect implementation.}

\subsection{$W$ state protocol}
\begin{figure}[!h]
	\centering
	\includegraphics[width=0.9\linewidth,keepaspectratio]{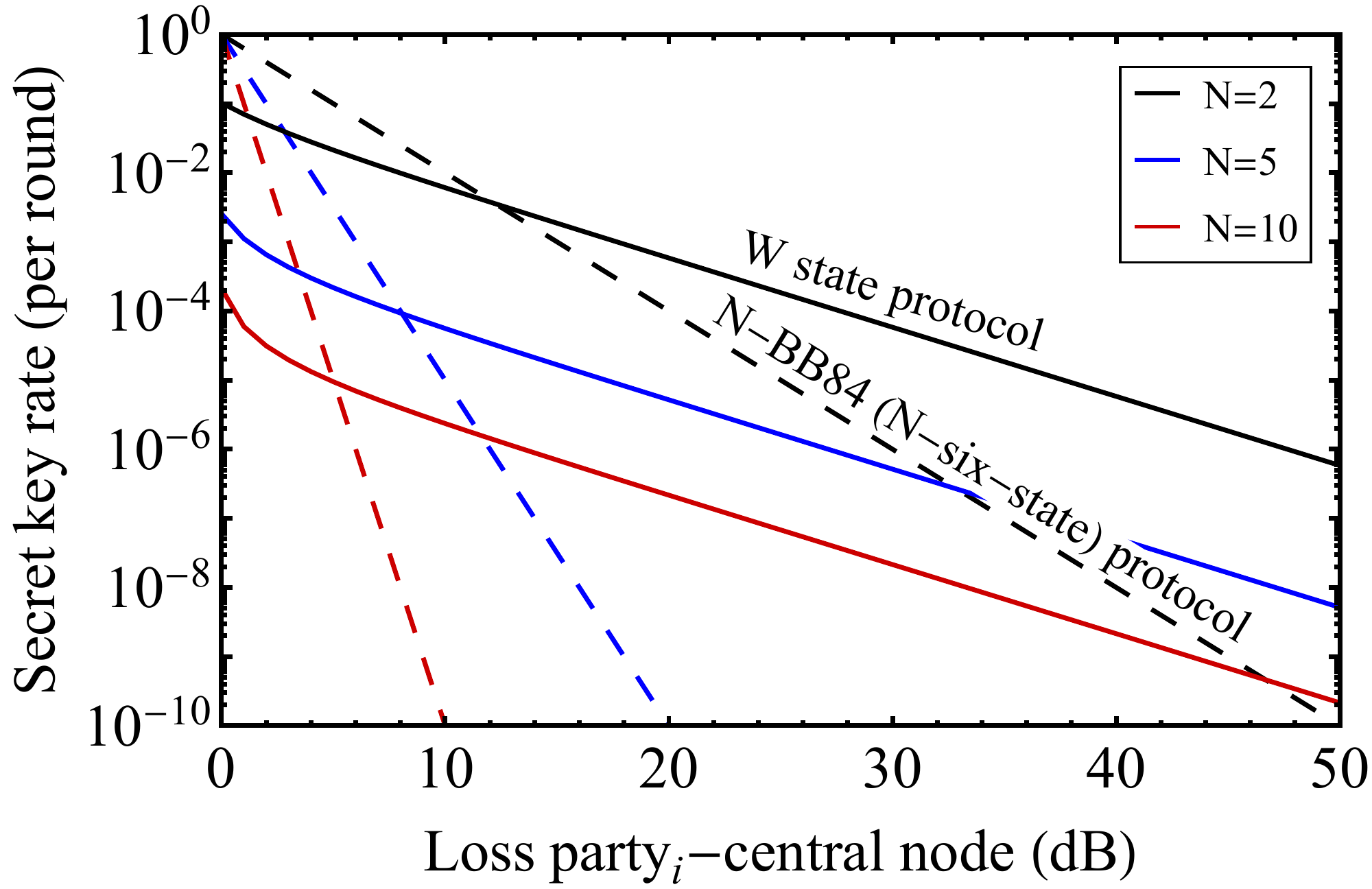}
	\caption{Comparison of the asymptotic conference key rate achieved by the W state protocol \cite{WstateProtocol} (solid) and by the N-BB84 protocol \cite{Grasselli_2018} (dashed, the N-six-state protocol rate is identical in this ideal scenario) as a function of the loss in the channel linking each party to the central entanglement distributor, for different number of parties ($N=2,5$ and $10$). We assume ideal implementations where the only source of error is photon loss and where the GHZ state of the N-BB84 (N-six-state) protocol is encoded in orthogonal polarizations of a photon.}
	\label{WstatevsBB84}
\end{figure}
\noindent
Quantum conference key agreement does not necessarily need to rely on the correlations provided by multipartite GHZ states. Indeed, the protocol devised in \cite{WstateProtocol} exploits the multipartite entanglement of a W-class state in order to establish a conference key. The W state of $N$ parties is defined as
\begin{align}\label{Wstate}
    \ket{{\rm W}_N}=\frac{1}{\sqrt{N}}\de{\ket{0\ldots01}+\ket{0\ldots10}+\ldots +\ket{1\ldots00}},
\end{align}
whereas a W-class state has a similar form to \eqref{Wstate} but presents arbitrary phases on each term.

In the conference key agreement protocol of Ref.~\cite{WstateProtocol}, the state is post-selected thanks to single-photon interference occurring in a central untrusted node, extending the founding idea of twin-field QKD \cite{TFLucamarini,TFCurty} to the multipartite scenario.

In particular, each round of the protocol starts with $\mathrm{party}_i$ ($i=1,2,\dots,N$) preparing the following entangled state between an optical pulse $a_i$ and a qubit $A_i$:
\begin{equation}
	\ket{\phi}_{A_i a_i} = \sqrt{q} \ket{0}_{A_i}\ket{0}_{a_i}+\sqrt{1-q} \ket{1}_{A_i} \ket{1}_{a_i} \,\,, \label{initial-state}
\end{equation}
where $\ket{0}_{a_i}$ is the vacuum state, $\ket{1}_{a_i}$ is the single-photon state, and $\{\ket{0}_{A_i},\ket{1}_{A_i} \}$ is the computational basis of the qubit. The state is strongly unbalanced towards the vacuum: $q\approx 1$. Every party sends his/her optical pulse to a central untrusted node through a lossy optical channel. Here, the pulses are combined in a balanced multiport beam splitter \cite{multiportBS} featuring a threshold detector at every output port. The central node announces whether each detector clicked or not and the parties only keep the rounds where exactly one detector clicked. These events are likely to be caused by the arrival and detection of just one photon, due to the unbalance towards the vacuum of the prepared state \eqref{initial-state}. Because of the balanced superposition generated by the multiport beam splitter, the detected photon could be sent by any party with equal probability. Thus, the main contribution to the $N$-qubit state shared by the parties conditioned on the single detection is a coherent superposition of states in which one qubit is in state $\ket{1}$ and all the others are in state $\ket{0}$, that is the mentioned $W$-class state. The qubits' relative coefficients have all equal weights but contain complex phases introduced by the multiport beam splitter.

It has been proven that the only multiqubit state yielding perfectly correlated and random outcomes upon performing local measurements is the GHZ state \cite{Epping}. Nevertheless, the post-selected $W$-class state can still be used to distil a conference key. More specifically, the parties obtain the key bits by measuring their qubit in a specific direction in the $X$-$Y$ plane of the Bloch sphere. The direction is the one that minimizes the bipartite QBER and depends on which detector clicked. For this reason, the protocol cannot be recast as a prepare-and-measure scheme, unlike its bipartite counterpart \cite{TFCurty}. Finally, the parties estimate the eavesdropper's knowledge by computing the expectation value of the $Z^{\otimes N}$ operator and by checking when it differs from the ideal case.
Note that if the parties are actually sharing a $W$-class state, then $\langle Z^{\otimes N}\rangle=-1$.

In \cite{WstateProtocol} the security of the protocol is proved in the finite-key regime and under coherent attacks performed by the eavesdropper.

The $W$-class $N$-qubit state on which the protocol is based is post-selected thanks to single-photon interference at the central node. Hence, the resulting key rate scales linearly with the transmittance $t$ of one of the quantum channels linking each party to the central node (if the channels are all symmetric). This contrasts with the honest implementations of the protocols \cite{Epping,Grasselli_2018} presented in subsection~\ref{sec:NBB84}, which are based on the distribution of $N$-qubit GHZ states. If these states are encoded, e.g., in the orthogonal polarizations of a photon, their key rate cannot scale better than $t^N$, where $t$ is the transmittance of the link between one party and the central distributor of the $N$-partite entangled state. This makes the protocol based on the $W$ state much more suited to high-loss scenarios than the protocols of subsection~\ref{sec:NBB84}. This is clear from Fig.~\ref{WstatevsBB84}, where we plot the asymptotic conference key rates of protocols \cite{WstateProtocol} (solid lines) and \cite{Epping,Grasselli_2018} (dashed lines) as a function of the loss in the quantum channel linking one party to the central node ($-10\log_{10} t$). We assume ideal implementations where photon loss is the only source of error. We observe the existence of a loss threshold above which the protocol based on the $W$ state \cite{WstateProtocol} outperforms the protocols based on the distribution of GHZ states \cite{Epping,Grasselli_2018}. Moreover, the required loss for which the protocol \cite{WstateProtocol} outperforms the protocols \cite{Epping,Grasselli_2018} decreases as the number of parties involved increases.

\section{Continuous variable conference key agreement}

\noindent
Quantum conference keys may also be established by means of continuous variable (CV) quantum systems. Following the first of such protocols \cite{FirstCVMDI}, which enables quantum conferencing among three parties without trusting the measurement devices, more general and refined protocols \cite{OLLP19,ZSG18} have been devised. The latter allow an arbitrary number of users to establish conference keys when linked to a central untrusted relay in a star network. These schemes would allow high-rate intra-city secure conferencing among several users.

Both protocols \cite{OLLP19,ZSG18} rely on the correlations generated by an $N$-mode CV GHZ state \cite{CVGHZstates}:
\begin{equation}
    \ket{\mathrm{CVGHZ}}_N=\frac{1}{\sqrt{\pi}} \int_{-\infty}^{\infty} dx \ket{x}^{\otimes N} \,\,, \label{CVGHZ}
\end{equation}
where $\{\ket{x}\}_x$ are the eigenstates of the $\hat{X}$ quadrature.
However, while in \cite{ZSG18} the central relay is required to generate such multipartite entangled state, in \cite{OLLP19} the state is post-selected thanks to a multipartite CV Bell detection at the central relay. In particular, in \cite{OLLP19} every user prepares a Gaussian-modulated coherent state $\ket{\alpha_k}$ ($k=1,\dots N$) and sends it to the central relay. Here, a suitable cascade of beam splitters followed by homodyne detections of either quadrature $\hat{X}$ or quadrature $\hat{P}$ implement the multipartite Bell detection, whose outcome is made public. The Bell detection projects the incoming coherent states onto the CV GHZ state \eqref{CVGHZ} up to displacements of the $N$ modes. By employing the public data of the Bell detection, the parties post-process the variables $\{\alpha_k\}_{k=1}^N$ describing the prepared coherent states and neutralize the effect of the displacements. They are thus left with variables whose correlations reproduce those of the original CV GHZ state \eqref{CVGHZ} and hence can be used to distil a conference key. This procedure closely resembles the seminal work on measurement-device-independent (MDI) QKD with discrete variables \cite{DVMDI}\gm{\cite{DVMDI-Pirandola}} and its CV counterpart \cite{CVMDI}, now applied to a multipartite scenario. Indeed, the fact that the measurements are only performed by the untrusted relay, makes the protocol in \cite{OLLP19} an MDI multipartite QKD protocol. Nevertheless, its performance does not decrease exponentially with the number of users since the CV Bell detection is a deterministic process, unlike its discrete-variable counterpart \gm{\cite{DVMDICKA3}}.

Note that, unlike the discrete-variable scenario, here the correlated variables $\{\alpha_k\}_{k=1}^N$ used to distil a binary key are complex numbers. Nevertheless, one can still express the resulting \gm{asymptotic key rate against collective attacks} in terms of their mutual information $I(\alpha_k,\alpha_{k'})$ \gm{\cite{CVQKDreview}} with the well-known Devetak-Winter formula \cite{DW}. 

Compared to \cite{OLLP19}, the protocol in \cite{ZSG18} is not MDI since the multipartite GHZ state generated in the untrusted relay is then distributed to the parties who perform trusted measurements. Moreover, from a practical point of view, this scheme is harder to implement, as it involves the preparation of several optical modes in squeezed states and their subsequent entanglement in a specific target state. Nevertheless, in principle, the scheme in \cite{ZSG18} could achieve slightly higher performances than the more practical protocol in \cite{OLLP19}.

In terms of security, both protocols \cite{OLLP19,ZSG18} have been proved to be secure against collective Gaussian attacks. Furthermore, the protocol in \cite{OLLP19} has been analyzed in the framework of finite-key composable security and proven to be secure against coherent attacks through a Gaussian de Finetti reduction \cite{GaussiandeFinetti}.

\section{Device-independent conference key agreement}\label{sec:DI}
\noindent
In the device-independent scenario, Alice and the Bobs do not want to assume any knowledge about the distributed system and internal working of their devices. 
\gm{Security can even be analyzed under the premise  that the shared states as well as the measurement devices were manufactured by the adversary.} \gm{We note that some assumptions are still present in the device-independent scenario, such as isolated labs and trusted random number generators (see~\cite{Murta19} for a discussion)}.
The parties' goal is to ensure security using only  the observed statistics of inputs and outputs. In a device-independent protocol security is certified by the violation of a Bell inequality. 

Note that in a device-independent conference key agreement (DICKA) protocol, an analysis against coherent attacks also needs to account for the fact that the eavesdropper might program the devices  to behave in different ways at each round of the protocol. In particular the measurement devices could have memory and behave in correlation with the outcomes of previous rounds. This makes the security analysis in the fully device-independent adversarial scenario significantly more intricate.

A recently developed technique \cite{EAT,EAT1}, the entropy accumulation theorem (EAT), provides the tools to perform the security analysis of device-independent protocols in the fully adversarial scenario maintaining some noise robustness. The EAT \cite{EAT,EAT1} extends the de Finnetti theorems \cite{KR05,CKR09} to the device-independent setting, allowing to reduce the analysis to collective attacks.

\begin{Rmk}(Composability in the device-independent scenario)\label{remarkDI}
The security definition, Definition~\ref{def:security}, implies universal composability of conference key agreement in the trusted device scenario.  However, for the device-independent scenario, attacks proposed in Ref.~\cite{BCK13} show that composability cannot be guaranteed if the same devices are re-used in a subsequent protocol. Indeed, in Ref.~\cite{BCK13} the authors describe attacks in which information about a previously generated key may be leaked through the public communication of a subsequent run of the protocol, if the devices are re-used. The attacks described in Ref.~\cite{BCK13} can be avoided if the parties have sufficient control of the internal memory of their devices and are able to re-set it after one execution of the protocol.
\end{Rmk}

\begin{figure}[!h]
	\centering
	\includegraphics[width=0.9\linewidth,keepaspectratio]{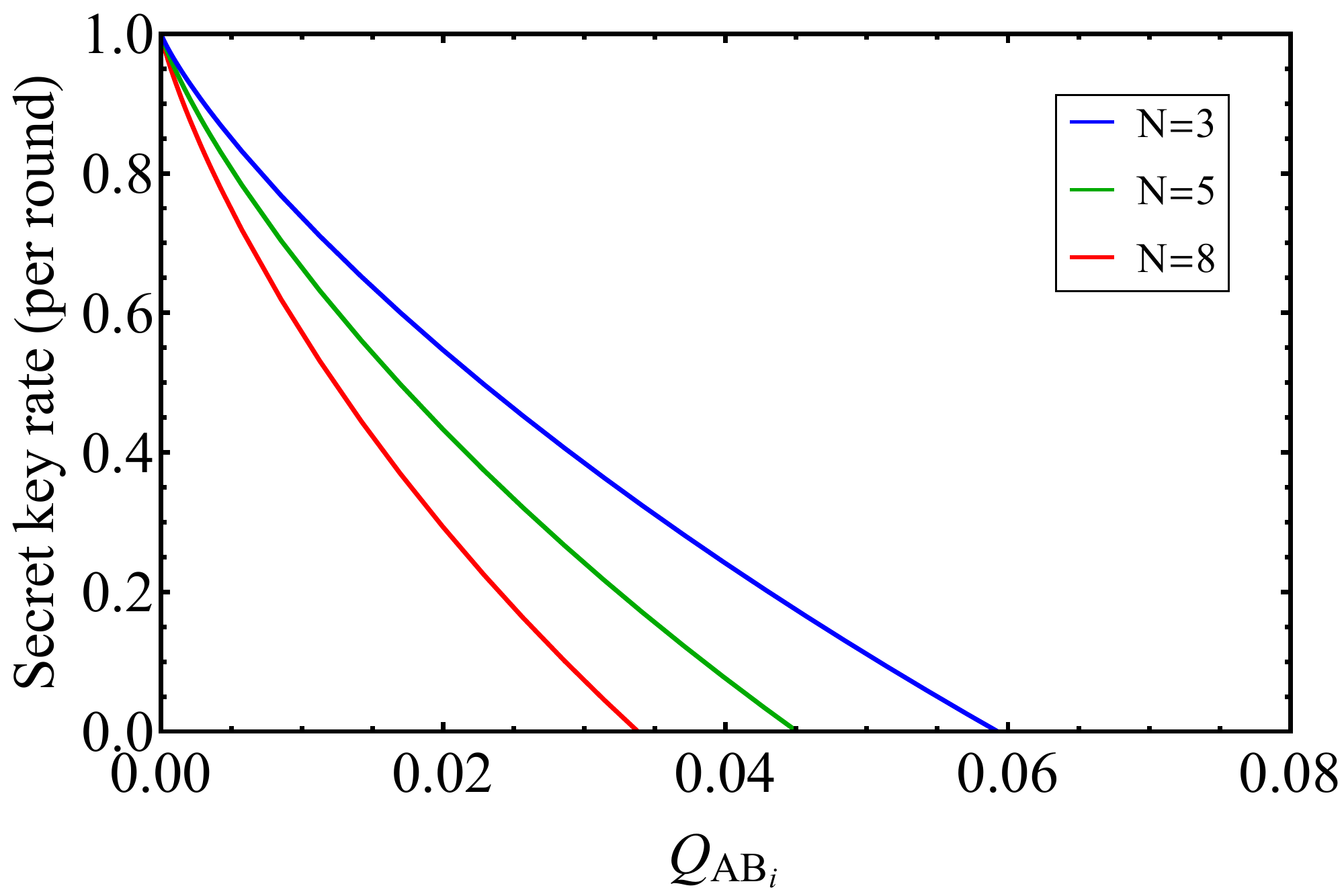}
	\caption{Asymptotic secret key rate for the DICKA protocol of Ref.~\cite{DICKAreply} as a function of the QBER and for fixed number of parties ($N=3,5,8$). We assumed an implementation where the $N$-party GHZ state is submitted to the depolarizing channel $\mathcal{D}_2^{\otimes N}(\ketbra{{\rm GHZ}_N}{{\rm GHZ}_N})$.}
	\label{di-plot}
\end{figure}

Based on the EAT, a DICKA protocol was proposed in \cite{DICKA,DICKAreply}.
The protocol of Ref.~\cite{DICKA} initially considers the multipartite Mermin-Ardehali-Belinskii-Klyshko (MABK) inequalities \cite{MABK1,MABK2,MABK3}. However, as shown in \cite{DICKAcomment}, the MABK inequalities are not suitable for establishing a conference key, as an overhead amount of information is required for information reconciliation. In Ref.~\cite{DICKAreply}, a new multiparty inequality is introduced and positive conference key can be established in the device-independent scenario.
{Fig. \ref{di-plot} shows the  asymptotic key rates for the device-independent protocol of Ref. \cite{DICKAreply} for $N=3, 5, 8$, for an implementation in which all the qubits are submitted to a depolarizing channel.}

The key rates derived in \cite{DICKAreply} are based on an analytical lower bound to von Neumann entropy of Alice's outcome conditioned on the information available to the eavesdropper, $H(A|E)$, as a function of the violation of the Bell inequality under consideration. The bound employs a relation between the considered multipartite inequality and the bipartite Clauser-Horne-Shimony-Holt (CHSH) inequality \cite{CHSH}.

In general, it is not possible to compute directly $H(A|E)$ as a function of the violation for an arbitrary Bell inequality. This is due to the lack of knowledge about the underlying system. A lower bound can be obtained using the relation $H(A|E)\geq H_{\min}(A|E)$, where $H_{\min}(A|E)$ is the conditional min-entropy {defined in \eqref{eq:Hmin}. Due to the relation with the guessing probability, \eqref{eq:Hmin},}
the conditional  min-entropy, $H_{\min}(A|E)$, can be estimated in the device-independent scenario \cite{MPA11} using the hierarchy of semi-definite approximations to the quantum set \cite{NPA1,NPA2}. This method is, however, computationally costly and may lead to non-tight bounds.

Bell inequalities tailored to DICKA protocols were further investigated in \cite{HKB19}, where the authors introduced a family of multipartite Bell inequalities {(containing the inequality of \cite{DICKAreply} as a special case)} that are maximally violated by the GHZ state, with the $Z$-basis being one of the optimal measurements for Alice. These are essential features to build a device-independent conference key agreement protocol.

It is interesting to remark that the MABK inequalities were previously explored in other multiparty communication protocols.
Refs.~\cite{SG01,SG_pra_01} consider a secret sharing scenario in which Alice distributes the key in such a way that the $N-1$ Bobs need to collaborate to retrieve its value.
The authors establish that, if the eavesdropper is restricted to individual attacks, then the violation of a MABK inequality can guarantee security, even if some of the Bobs collaborate with Eve. Even though this scenario was initially denoted $N$-party QKD \cite{SG01,SG_pra_01}, it should be distiguished from the scenario we consider in this review: in which the goal is that all the Bobs can retrieve the key independently.

\section{Multipartite private states}
\noindent
Most of the quantum conference key agreement protocols presented in the previous sections  exploit the correlations of the multipartite GHZ state \eqref{GHZstate}. Therefore, GHZ distillation protocols are in close connection with distillation of secret conference keys. Indeed, if the parties share several copies of a resource state that can be turned into a smaller number of GHZ states, then they could perform a distillation protocol followed by measurements to generate a secret key. The connection of entanglement distillation and conference key agreement protocols is discussed in \cite{CL07}.

However, it is not only through distillation of GHZ states that one can obtain a secret key. Indeed, as shown in ~\cite{AH09}, an $\epsilon$-secure conference key can also be obtained from bound entangled states. This result generalizes an analogous one derived in the bipartite case \cite{HHHO05}.

The concept of private states \cite{HHHO09} was generalized to the multipartite case in Ref.~\cite{HA06,AH09}.
Similar to the bipartite case, a multipartite private state can be seen as a twisted GHZ state tensored with an extra density matrix (the shield)
\begin{align}\label{privatstate}
    \Gamma_{A\vec{B}A'}^{(d)}=U_t(\ketbra{{\rm GHZ}_N^d}{{\rm GHZ}_N^d}\otimes \rho_{A'})U_t^{\dag},
\end{align}
where $\ket{{\rm GHZ}_N^d}=\frac{1}{\sqrt{d}}\sum_{i=0}^{d-1}\ket{ii\ldots i}$ is the $N$-party GHZ state of dimension $d$ and the multipartite twisting is a unitary operation of the form
\begin{align}
    U_t=\sum_{i_1,\ldots,i_N=0}^{d-1}\ketbra{i_1\ldots i_N}{i_1\ldots i_N}\otimes U_{i_1,\ldots,i_N}
\end{align}
for arbitrary unitaries $U_{i_1,\ldots,i_N}$ acting on $A'$.

Ref.~\cite{AH09} establishes that if from a resource state Alice and the Bobs can distill an $\epsilon$-secret conference key, then there exists an LOCC protocol that can distill a state close to a private state \eqref{privatstate} and vice-versa. They also exhibit examples of multipartite bound entangled states, that are states from which a GHZ state cannot be distilled, which are $\epsilon$-close to private states. This establishes that distillation of GHZ states is not necessary for quantum conference key agreement and more general classes of protocols are possible.
\gm{Limits on the performance of private states distribution in a network, with and without quantum repeaters, and its consequence for CKA protocols, has recently been investigated in~\cite{DBWH19,Pirandola19}.}

{In the framework of quantum channels and private state distillation 
using multiplex channels, 
Ref.~\cite{DBWH19} establishes that genuine multipartite entanglement is necessary for single shot key distillation.}
This implies that, if a key can be distilled from $n$ copies of a multipartite state $\rho$, then $\rho^{\otimes n}$ needs to be genuine multipartite entangled. However, this does not require that genuine multipartite entanglement is present at the single round level $\rho$. Indeed, a study of the entanglement properties required for a resource state to enable a conference key was recently performed in~\cite{Giacomo}. Results of Ref.~\cite{Giacomo} show that a conference key can be established even if the parties 
{share a biseparable state in every round.}

\section{Outlook}
\noindent
We reviewed the state-of-the-art quantum CKA schemes based on multipartite entanglement. We discussed proposed protocols and their security proofs under different levels of assumptions for the characterisation of the devices, and for several types of implementations. 

From an experimental point of view, the implementation of quantum CKA is increasingly accessible, due to key developments of its fundamental ingredients.
Multipartite entanglement has been generated in a variety of physical systems, such as e.g.\ ion traps \cite{Leibfried2005,Haffner2005,blatt2011}, photonic systems \cite{pan2012,photon12,photon16,photon18,Malik2016}, superconducting circuits \cite{superconducting2014,superconducting2017,superconducting2019} and nuclear spin qubits in diamond \cite{vandam2019}.
Also, entanglement among several particles is naturally generated in atomic ensembles~\cite{AtomicEnsembleReview1,AtomicEnsembleReview2}, and methods to quantify and manipulate this entanglement are being developed~\cite{Lucke14,Riedel2010,Kunkel413,Fadel409,Lange416}.
\gm{Even a thermalised interacting photon gas \cite{Weitz2017} has shown potential to be a source of genuine
multipartite entanglement.}
Recently, the first quantum CKA protocol has been implemented~\cite{CKAexperiment} among four parties. The experiment is based on the multiparty BB84 protocol \cite{Grasselli_2018} discussed in Section~\ref{sec:NBB84}. It relies on the generation of polarization-encoded four-party GHZ states at telecom wavelength by a central quantum server. The states are then distributed to the four parties over up to 50~km of optical fibers, generating a secure conference key according to Definition~\ref{def:security}.

While experimental progress is still necessary to scale implementations of quantum CKA to many users, improvements from the theory side are crucial to reduce the experimental demands. To this aim, the development of new protocols and new techniques to prove security will contribute to make quantum CKA a feasible technology.

Novel protocols exploring different resource states and network architectures can lead to improved performance and noise robustness.
In the bi-partite case, QKD protocols for $d$-dimensional systems achieve higher rates and better noise tolerance~\cite{dQKD} than the qubit-based protocols. In order to explore this possibility in the multipartite case, quantum CKA protocols for $d$-dimensional systems need to be developed.
Such a generalization can also find applications in the layered protocol presented in~\cite{layeredQKD}. In Ref.~\cite{layeredQKD}, asymmetric high-dimensional multipartite entangled states are used to design a layered protocol that establishes a secret key simultaneously between different subsets of users in a network. 

Similarly, new tools to improve security proofs can lead to better rates and noise tolerance, especially for DICKA protocols. A family of Bell inequalities suitable for conference key agreement protocols has been introduced in~\cite{HKB19}. However, only non-tight numerical lower bounds to the key rates are currently available for DICKA protocols based on these inequalities. The introduction of tighter analytical bounds addressing their security proofs could lead to higher key rates in DICKA protocols.

\section*{Acknowledgements}
\noindent
We thank Giacomo Carrara, Timo Holz, Carlo Ottaviani, Stefano Pirandola, {Siddhartha Das, Stefan B\"auml, Marek Winczewski, and Karol Horodecki}  for helpful discussions. This work was funded by the Deutsche Forschungsgemeinschaft (DFG, German Research
Foundation) under Germany's Excellence Strategy - Cluster of Excellence
Matter and Light for Quantum Computing (ML4Q) EXC 2004/1 - 390534769, by the European Union’s Horizon 2020 research and innovation programme under the Marie Sk{\l}odowska-Curie grant agreement No~675662, and
by the Federal Ministry of Education and Research BMBF (Project Q.Link.X and HQS).

\bibliographystyle{ieeetr}
\bibliography{biblioCKA}

\end{document}